# Angle-resolved hollow-core fiber-based curvature sensor

William M. Guimarães [1], Cristiano M. B. Cordeiro [2, *], Marcos A. R. Franco [1], Jonas H. Osório [2]

[1] Institute for Advanced Studies, IEAv, São José dos Campos, Brazil
[2] Institute of Physics "Gleb Wataghin", University of Campinas, UNICAMP, Campinas, Brazil
* cmbc@ifi.unicamp.br

**Abstract:** We propose and theoretically study a new hollow-core fiber-based curvature sensor with the capability of detecting both bending magnitude and angle (direction). The new sensor relies on a tubular-lattice fiber that encompasses, in its microstructure, tubes with three different thicknesses. By adequately choosing the placement of the tubes within the fiber cross-section, and by exploring the spectral shifts of the fiber transmitted spectrum due to the curvature-induced mode field distributions' displacements, we demonstrate a multi-axis bending sensor. In the proposed sensor, curvature radii and angles are retrieved via a suitable calibration routine, which is based on conveniently adjusting empirical functions to the sensor response. Evaluation of the sensor performance for selected cases allowed to determine the curvature radii and angles with percentual errors of less than 7%. The approach proposed herein provides a promising path for the accomplishment of new curvature sensors able to resolve both the curvature magnitude and angle.

## 1. Introduction

The great and growing interest in hollow-core photonic crystal fibers (HCPCF) technology motivates intense research efforts by the photonics community. The prospects of their applications in both fundamental and applied fields allow them to be identified as a truly enabling technology that will provide a new scenery for next-generation optical devices [1]. Indeed, such a new framework for the developments of the next-generation HCPCF-based devices is provided by the recent endeavors on the optimization of both the HCPCF designs and fabrication methods which, in turn, have entailed an expressive reduction of the attenuation values in the infrared [2-4], visible and ultraviolet ranges [5-7].

Among the application opportunities, which encompass, for example, the development of novel optical sources and atom optics experiments [8-10], the field of optical sensors acquires a prominent position within HCPCF technology due to the broad set of parameters that can be potentially monitored. HCPCFs have been demonstrated, for example, to be a promising platform for probing concentrations of chemical species in gaseous and liquid samples [11-16]. Additionally, they have been employed in gyroscopes [17, 18] and strain sensing experiments [19]. Very recently, we proposed a new single-ring tubular-lattice (SR-TL) HCPCF whose microstructure comprises a ring of tubes with five different thicknesses and demonstrated that its transmission characteristics are dependent on the applied curvature and its direction [20]. This new fiber structure, together with other technologies such as off-centered core fibers [21] and multicore fibers [22], represent promising opportunities for the realization of directional curvature sensing. Moreover, this novel structure identifies a new avenue for the utilization of HCPCF with structural asymmetries, formerly employed to explore birefringent and polarizing HCPCFs [23, 24], as well as to alter the transmission loss hierarchy of the modes guided through the fiber core [25].

In this context, we here simplify our recently reported SR-TL HCPCF by proposing and theoretically studying a new fiber structure that encompasses a ring of tubes with three different thicknesses. Having three different thicknesses within the fiber structure (instead of five, as previously proposed in [20]), identifies a significant simplification path considering the fabrication viability of the fiber and, therefore, represents an important advancement concerning its future experimental realization.

In the fiber proposed herein, by judiciously choosing the positioning of the tubes with different thicknesses within the fiber microstructure, we obtain an azimuthally asymmetric HCPCF endowed with a bend-sensitive and angle-resolved (direction-resolved) transmission response. Indeed, the dependence of the fiber transmission characteristics on the curvature radii and angles is attributed to the bend-induced displacements of guided-mode field distribution within the fiber core [20]. Additionally, as the core of the fiber reported here sits at the fiber geometrical center, its integration with standard optical fibers could be potentially simplified compared to multicore and surface-core fiber-based directional curvature sensors [21, 22]. Indeed, current HCPCF technology allows the realization of high-quality splices between HCPCFs and standard solid-core optical fibers, as recently reported in the literature [26, 27].

To study the sensing capabilities of the proposed fiber, we report on a calibration routine based on conveniently fitting the spectral shifts of the transmission spectra for different curvature radii and angles. By employing our methods, we demonstrate a multi-axis HCPCF-based curvature sensor, able to retrieve both bending radius and angle. We understand that our research provides a further step to the assessment of the bend-dependent response of azimuthally asymmetric HCPCFs and a promising path for the accomplishment of novel direction-resolved curvature sensors.

The manuscript is organized as follows. We start by exposing the proposed fiber structure and its design rationale. In sequence, we provide details on the simulation setup and the conventions which have been adopted to study the sensor characteristics. Finally, we investigate the fiber's bend-dependent response, describe the sensor calibration routine, and perform tests to demonstrate the sensor's operation and effectiveness.

## 2. Materials and methods

Figure 1a presents a diagram of the fiber structure we propose herein. It consists of an SR-TL HCPCF whose cladding displays tubes with three different thicknesses, $t_0$, $t_1$, and $t_2$. The tubes with thicknesses $t_1$ and $t_2$ are placed in orthogonal directions within the fiber microstructure. Such configuration of cladding tubes allows obtaining a bend-dependent fiber transmission spectrum and attaining an angle-resolved curvature sensor, as one will detail in the following. The curvature radius is defined as $R$ and the curvature angle (direction) is represented by $\theta$ (as illustrated in Figure 1a). In turn, $D_C$ represents the core diameter and $D_t$ the diameter of the cladding tubes. Under our convention, $\theta = 0°$ refers to the situation in which the center of the curvature is located in the +x direction, and $\theta = 90°$ stands for a curvature whose center lies in the +y direction. Analogously, $\theta = 180°$ refers to a bend towards –x direction and, and $\theta = 270°$ means a bend towards the –y direction.

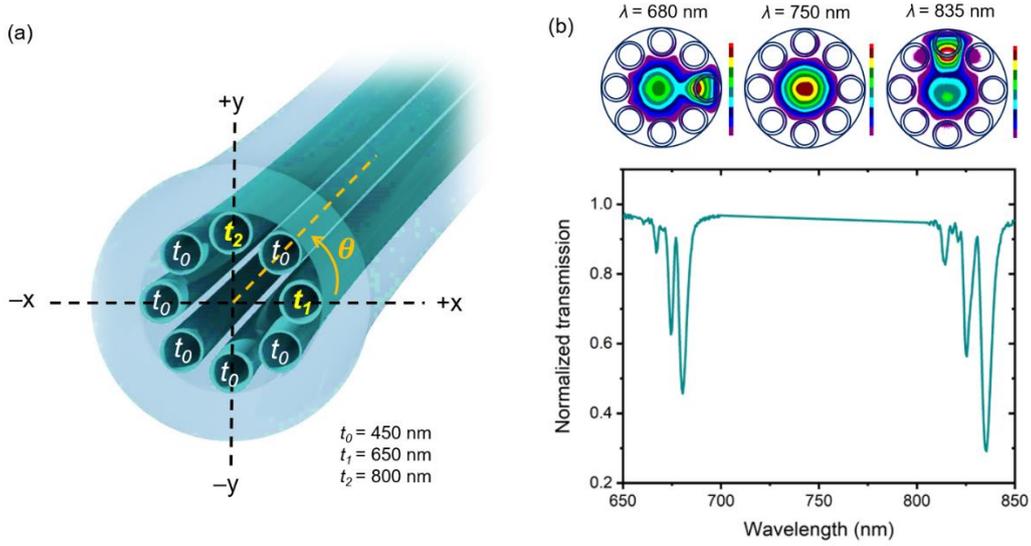

**Figure 1.** (a) HCPCF cross-section diagram ($t_0$, $t_1$, and $t_2$: thicknesses of the cladding tubes; $\theta$: curvature angle). (b) Typical transmission spectrum for a straight fiber and mode profiles at selected wavelengths.

Guidance of light in SR-TL HCPCF relies on the minimization of the coupling between core and cladding modes [28, 29]. At specific wavelengths, strong coupling between the core mode and the cladding tubes' modes occurs, and high-loss spectral intervals are observed in the fiber transmission spectrum. Such high-loss wavelengths (resonances), $\lambda_m$, are described by, $\lambda_m = (2t/m)\sqrt{n_2^2 - n_1^2}$, where $t$ is the thickness of the cladding tubes, $n_1$ is the refractive index of the core, $n_2$ is the refractive index of the cladding material, and $m$ is the resonance order. Here, we consider $n_1$ = 1 (for air) and $n_2$ = 1.45 (for silica). The value of $n_2$ = 1.45 has been used in the simulations for simplicity as, within the wavelength range considered in the simulations (from 650 nm to 850 nm), the refractive index of silica varies from 1.4565 to 1.4525 [30]. The thickness of the cladding tubes determines the resonant wavelengths and, hence, the spectral positions of the high loss regions in the fiber transmission spectrum. In the fiber proposed herein, as the tubes in the fiber structure display three different thicknesses, several resonances are expected to appear in the fiber transmission spectrum.

Figure 1b presents a typical transmission spectrum of the fiber structure represented in Fig. 1a when $t_0$ = 450 nm, $t_1$ = 650 nm, and $t_2$ = 800 nm (for a straight fiber). Here, it is worth observing that the choice of the thicknesses of the cladding tubes is constrained to two aspects. The first one considers that the thicknesses of the tubes must be such to provide sufficiently separated resonances in the fiber transmission spectrum (so it does not hinder the identification of the resonances' shifts when performing the sensing studies). The second aspect considers the fabrication tolerances in state-of-the-art HCPCF (~1% cross-sectional tube thickness variation [31]). The thicknesses $t_0$, $t_1$, and $t_2$ used in the simulations reported herein meet these two constraints.

The simulation has been carried out by using the Beam Propagation Method (BPM) and by considering a fiber length of 1 mm. Such fiber length, to be maintained in all the simulations in this manuscript, has been chosen as a good compromise between the computational cost for the simulations and, still, to be reasonably large to allow adequate observation of the high loss regions in the fiber transmission spectrum and suitable characterization the optical response of the fiber. In this context, a plausible path towards a practical realization of the sensor would be to splice a millimeter-long HCPCF section in-between standard solid-core optical fibers, as this method would allow to suitably assess the high loss transmission spectral regions and to apply the curvature conditions to the fiber. Indeed, the routines for splicing short fiber lengths are well established in multimode interference devices technology [32, 33] and could be customized for HCPCF. Additionally, the recent results on high-quality HCPCF splices [26, 27] demonstrate that the fiber proposed in this manuscript can be adequately integrated with standard solid-core optical fibers.

Insets in Fig. 1b display the simulated intensity distributions at the fiber output at selected wavelengths. Around $\lambda$ = 680 nm, coupling between the core mode and the modes in the tube with thickness $t_1$ occurs (tube at +x direction). Analogously, around $\lambda$ = 835 nm, coupling between the core mode and the modes in the tube with thickness $t_2$ takes place (tube at +y direction). Around $\lambda$ = 750 nm, the guided mode is highly confined into the fiber core, as, at this wavelength, robust inhibition of the coupling between the core and cladding modes is achieved.

## 3. Results

### 3.1. The fiber's bend dependent response

As demonstrated in our previous work [20], when the fiber is bent, the curvature-induced mode field displacements entail blueshifts or redshifts in the resonances' spectral positions. Indeed, the shifting of the resonances' wavelengths is determined by the displacement trend of the mode field distribution inside the fiber core due to the bending. Here, we recall that, as studied in [20], bending the fiber causes the intensity distribution of the core-guided modes to shift towards the outer side of the curvature, as typically observed in SR-TL HCPCF [34, 35].

Figure 2 presents the simulated fiber transmission spectra around $\lambda$ = 680 nm and $\lambda$ = 835 nm (*i.e.*, around the resonances associated to $t_1$ and $t_2$, respectively) for different curvature radii and representative curvature angles $\theta$. In Figure 2, we observe that, depending on the curvature direction and radius, the resonance spectral positions can be shifted towards shorter or longer wavelengths, as referenced in the latter paragraph. For example, when the center of curvature lies on the +x direction ($\theta$ = 0º), the guided mode intensity distribution is displaced away from the tube with thickness $t_1$ and the resonance around 680 nm blueshifts for smaller $R$ (Figure 2a). Otherwise, when the fiber is bent towards the –x direction ($\theta$ = 180º), the guided mode intensity distribution is displaced towards the tube with thickness $t_1$, and the resonance around 680 nm redshifts for smaller $R$ (Figure 2b).

An analog behavior is observed for the spectral shifts associated with the resonances of the tube with thickness $t_2$. When the center of curvature lies on the +y direction ($\theta$ = 90º), the guided mode intensity distribution is displaced away from the tube with thickness $t_2$ and its correspondent resonance blueshifts when $R$ decreases (Figure 2c). Otherwise, if the fiber is bent towards the –y direction ($\theta$ = 270º), the guided mode intensity distribution is displaced towards the tube with thickness $t_2$ and the resonance around 835 nm redshifts for reduced $R$ (Figure 2d). Remarkably, one observes virtually no impact on the spectral positions of the resonances when the bending direction is orthogonal to the tubes' location within the fiber microstructure (*i.e.*, $t_1$ resonance does not shift for $\theta$ = 0º and $\theta$ = 180º, and $t_2$ resonance does not shift for $\theta$ = 90º and $\theta$ = 270º).

Additionally, it is worth mentioning that altering the curvature angle for a fixed curvature radius implies shifting of the spectral positions of the tubes' resonances. To exemplify such a situation, we show, in Figure 3, the fiber transmission spectra around the resonance associated with $t_1$ and $t_2$ for a fixed curvature radius ($R$ = 7 cm) and representative curvature angles.

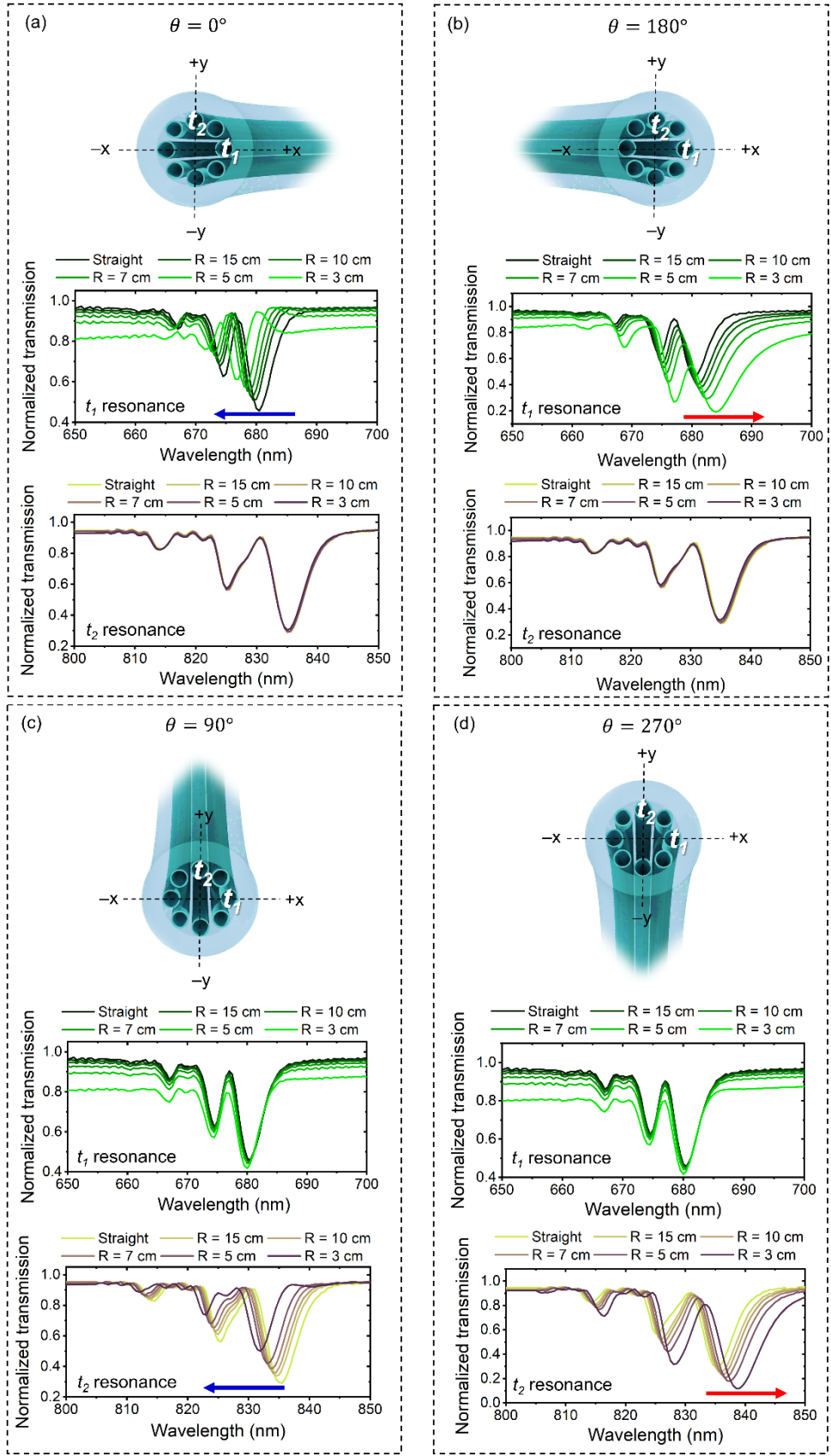

**Figure 2.** Simulated transmission spectra within the wavelength ranges associated to $t_1$ and $t_2$ resonances for different curvature radii ($R$) and angles (a) $\theta$ = 0º, (b) $\theta$ = 90º, (a) $\theta$ = 180º, and (b) $\theta$ = 270º. The blue and red arrows identify, respectively, the blueshifts and redshifts of the resonances due to bending.

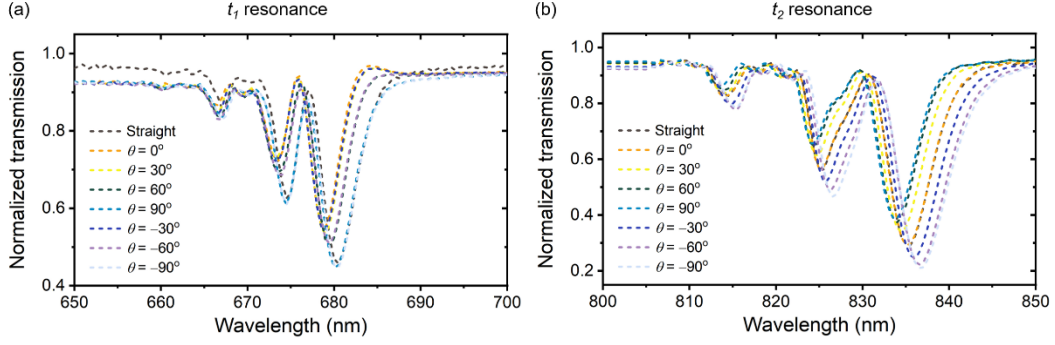

**Figure 3.** Transmission spectra around the (a) $t_1$ and (b) $t_2$ tube resonances for R = 7 cm and representative curvature angles, $\theta$.

*3.2. Sensor calibration*

Figure 4 consolidates the spectral shifts of the resonances around $\lambda_1$ = 680 nm and $\lambda_2$ = 835 nm (identified as $\Delta\lambda_1$ and $\Delta\lambda_2$, respectively) for $\theta$ between –180° and 180° and R = 3 cm, 5 cm, 7 cm, 10 cm, and 15 cm. The spectral shifts presented in Figure 4 are defined as $\Delta\lambda \equiv \lambda_{R,\theta} - \lambda_{R=0}$, where is $\lambda_{R,\theta}$ stands for the resonant wavelength for a fiber bent at a curvature radius $R$ and angle $\theta$, and $\lambda_{R=0}$ denotes the resonant wavelength for the straight fiber. In a hypothetical experimental realization of the sensor, $\Delta\lambda$ could be determined by coupling broadband light into the fiber core and by following the resonances' wavelength shifts in the fiber transmission spectrum.

Data in Figure 4 allow calibrating the sensor optical response by conveniently fitting $\Delta\lambda$ for different $R$ and $\theta$. Here, we empirically choose a sinusoidal fitting function, as shown in Equation 1, to account for $\Delta\lambda$. The choice of the function in Equation 1 has been driven by the observation of the sinusoidal trend of the data points in Figure 4. In Equation 1, $\Delta\lambda_0$, $\Lambda$, $\theta_0$, and $\delta$ are fitting parameters. The fitted functions are presented as dashed lines in Figure 4. The coefficients of determination ($R^2$) associated with the fits range from 0.987 to 0.999 for $t_1$ resonance and from 0.972 to 0.977 for $t_2$ resonance.

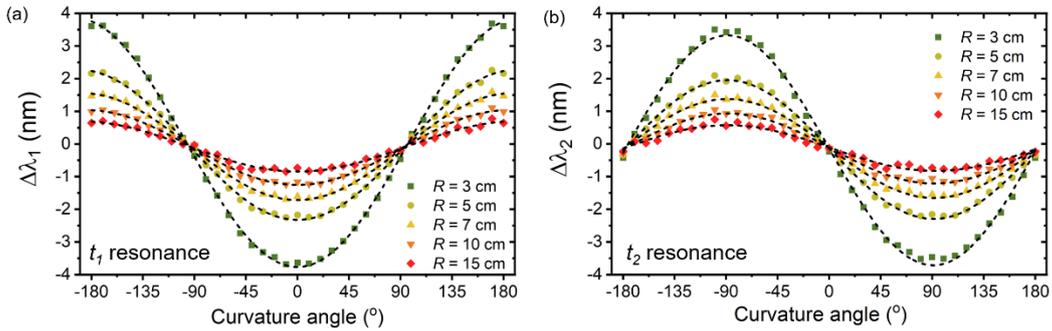

**Figure 4.** Wavelength shifts as a function of the curvature angle ($\theta$) for different curvature radii ($R$) for the resonances associated with the tubes with thicknesses (a) $t_1$ ($\Delta\lambda_1$) and (b) $t_2$ ($\Delta\lambda_2$).

$$\Delta\lambda = \Delta\lambda_0 + \Lambda \sin\left[\frac{\pi(\theta-\theta_0)}{\delta}\right] \tag{1}$$

To attain a suitable calibration of the fiber optical response, the parameters $\Delta\lambda_0$, $\Lambda$, $\theta_0$, and $\delta$ can be plotted as a function of $R$ and fitted by using empirical functions. Figure 5 exhibits graphs of $\Delta\lambda_0$, $\Lambda$, $\theta_0$, and $\delta$ as a function of $R$, together with the fitting curves. Table 1 informs the fitting functions which have been used to adjust $\Delta\lambda_0$, $\Lambda$, $\theta_0$, and $\delta$ trends as a function of $R$. It is worth observing that $\Lambda$ is related to the amplitudes of the wavelength shifts and $\delta$ to the conversion between radians and degrees. In turn, $\theta_0$ allows accounting for the cosine-like and sine-like behaviors of $\Delta\lambda_1$ and $\Delta\lambda_2$, respectively.

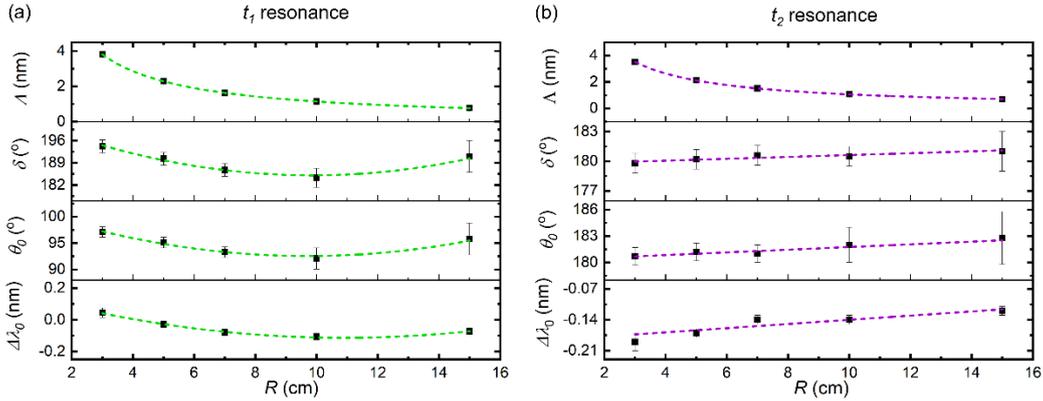

**Figure 5.** Parameters $\Delta\lambda_0$, $\theta_0$, $\delta$ and $\Lambda$ as a function of the curvature radius ($R$), corresponding to the resonances associated to the tubes with thicknesses (a) $t_1$ and (b) $t_2$.

**Table 1.** Fitting functions used to adjust $\Delta\lambda_0$, $\theta_0$, $\delta$ and $\Lambda$ trends as a function of $R$ ($a$, $b$, $c$, and $d$ are representative fitting constants).

| Parameter | Fitting function ($t_1$ resonance) | Fitting function ($t_2$ resonance) |
| --- | --- | --- |
| $\Lambda$ | $\Lambda = a/R^b$ | $\Lambda = a/R^b$ |
| $\delta$ | $\delta = a + bR + cR^2$ | $\delta = a + bR$ |
| $\theta_0$ | $\theta_0 = a + bR + cR^2$ | $\theta_0 = a + bR$ |
| $\Delta\lambda_0$ | $\Delta\lambda_0 = a + bR + cR^2$ | $\Delta\lambda_0 = a + bR$ |

As the dependency of $\Delta\lambda_0$, $\Lambda$, $\theta_0$, and $\delta$ with $R$ has been obtained by the fitting curves in Figure 5, one can assume $\Delta\lambda_0$, $\Lambda$, $\theta_0$, and $\delta$ as functions of $R$ and, thus, rewrite Equation 1 by considering $\Delta\lambda$ as a function of $R$ and $\theta$. We obtain, therefore, Equation 2, where $\Delta\lambda_0(R)$, $\Lambda(R)$, $\theta_0(R)$, and $\delta(R)$ are the fitted functions.

$$\Delta\lambda(R,\theta) = \Delta\lambda_0(R) + \Lambda(R)\sin\left[\frac{\pi(\theta - \theta_0(R))}{\delta(R)}\right] \quad (2)$$

Figure 6 shows the 3D-plots for $\Delta\lambda(R,\theta)$, associated to the resonances around $\lambda_1 = 680$ nm and $\lambda_2 = 835$ nm – $\Delta\lambda_1(R,\theta)$ and $\Delta\lambda_2(R,\theta)$, respectively –, which have been calculated by considering Equation 2 and the adjusted functions. Indeed, these plots are the basis of the sensing operation principle reported herein, which relies on accounting for the pair ($\Delta\lambda_1$, $\Delta\lambda_2$) and associating it to the bending radius and angle univocally. The analytical functions shown in Figure 6 are, thus, the calibration curves of the sensor. Indeed, the plots in Figure 6 are strongly correlated to the ones shown in Figure 4, from which the fitting parameters have been determined.

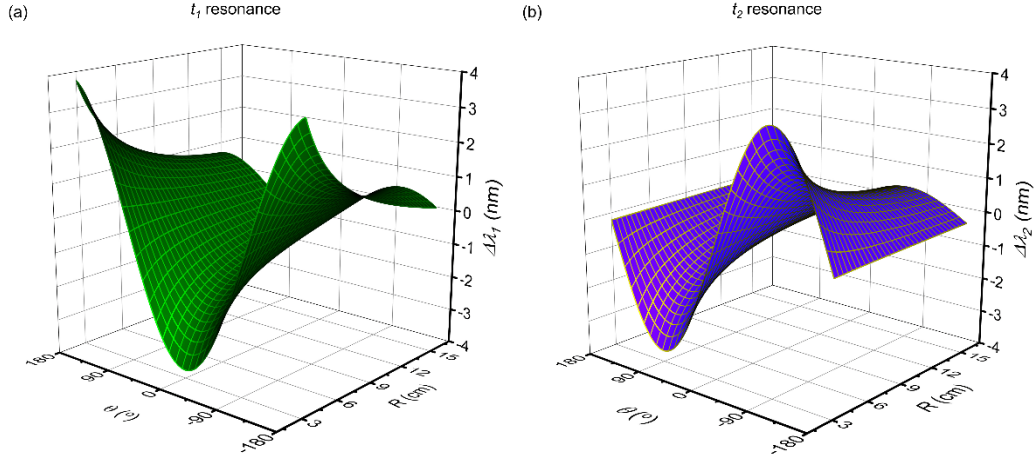

**Figure 6.** Sensor calibration curves, (a) $\Delta\lambda_1(R,\theta)$ and (b) $\Delta\lambda_2(R,\theta)$.

## 4. Discussion

To demonstrate the sensor's operation and its effectiveness, we study a set of situations by comparing, for selected values of $R$ and $\theta$, the corresponding wavelength shifts that are calculated from the BPM simulations and the ones that are accounted from the fitted functions. In the studied cases, we considered $R$ and $\theta$ values which are different from the ones employed in the system calibration fittings, but which are within the sensor calibration interval, *i.e.*, 3 cm < $R$ < 15 cm and -180° < $\theta$ < 180°.

Figure 7 illustrates the method for obtaining $R$ and $\theta$ from $\Delta\lambda_1$ and $\Delta\lambda_2$. The first studied case (Test #1, Figure 7a) considers that, in a hypothetical measurement, $\Delta\lambda_1$ and $\Delta\lambda_2$ have been measured as $\Delta\lambda_1$ = (2.48 ± 0.05) nm and $\Delta\lambda_2$ = (−1.45 ± 0.05) nm. By intersecting the latter $\Delta\lambda_1$ and $\Delta\lambda_2$ values with system calibration curves (as shown in Figure 6), we obtain the 2D color plots shown in Figure 7a, which maps the $R$ and $\theta$ that could correspond to the considered $\Delta\lambda_1$ and $\Delta\lambda_2$. The intersecting region between $\Delta\lambda_1$ and $\Delta\lambda_2$ plots readily communicates the $R$ and $\theta$ values expected from the sensor calibration. For the $\Delta\lambda_1$ and $\Delta\lambda_2$ values used in this example, the sensor calibration allows determining $R$ = (4.1 ± 0.2) cm and $\theta$ = (30 ± 2)°. Indeed, BPM simulations using $R$ = 4.0 cm and $\theta$ = 30° yields $\Delta\lambda_1$ = −2.48 nm and $\Delta\lambda_2$ = −1.45 nm. The method reported herein allow, therefore, obtaining $R$ and $\theta$ via the knowledge on $\Delta\lambda_1$ and $\Delta\lambda_2$.

Figure 7 also shows three other examples of the sensor operation. In Figure 7b, we assume that a hypothetical measurement allowed determining $\Delta\lambda_1$ = (−1.72 ± 0.05) nm and $\Delta\lambda_2$ = (−1.00 ± 0.05) nm (Test #2). Similarly, Figure 7b presents color maps on the considered $\Delta\lambda_1$ and $\Delta\lambda_2$ values as accounted from the sensor calibration. In this case, our methods allow determining $R$ = (6.2 ± 0.3) cm and $\theta$ = (30 ± 2)°, which are, once again, in good agreement with BPM simulations results (which yields $\Delta\lambda_1$ = −1.72 nm and $\Delta\lambda_2$ = −1.00 nm for $R$ = 6.0 cm and $\theta$ = 30°). The two other examples' results (Test #3 and Test #4) are summarized in Table 2 (together with the data from Test #1 and Test #2). Remarkably, the sensor reported herein allows determining the curvature radius and its direction if $\Delta\lambda_1$ and $\Delta\lambda_2$ are known. All the $R$ and $\theta$ values retrieved in the test cases are consistent with the estimated error bars and the maximum percentual error between expected and retrieved values have been found as 5.8% for $R$ values and 6.6% for $\theta$ values. A new angle-resolved curvature sensor based on HCPCF technology has thus been developed.

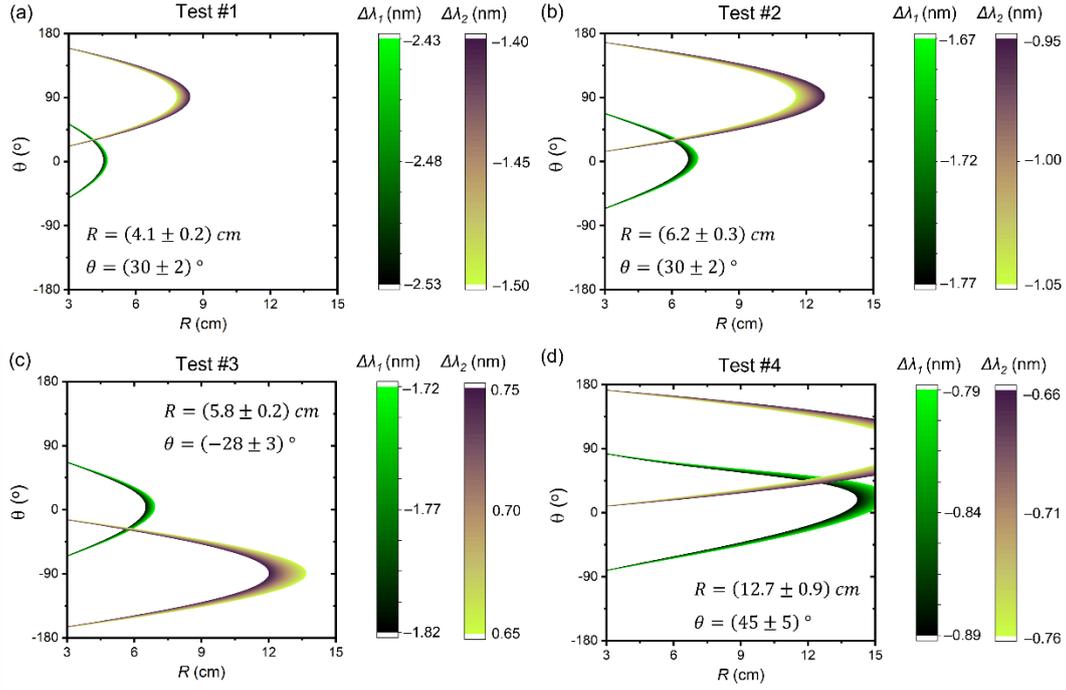

**Figure 7.** Color plots accounted from sensor calibration for selected $\Delta\lambda_1$ and $\Delta\lambda_2$ values. The intersecting region between $\Delta\lambda_1$ and $\Delta\lambda_2$ plots allows obtaining the curvature radius and angle expected from the sensor calibration.

**Table 2.** Summary of the sensor operation tests.

| Test | Results accounted from the system calibration | Results accounted from BPM simulations |
|---|---|---|
| #1 (Figure 7a) | $\Delta\lambda_1 = (-2.48 \pm 0.05)\ nm$<br>$\Delta\lambda_2 = (-1.45 \pm 0.05)\ nm$<br>$\Leftrightarrow$<br>$R = (4.1 \pm 0.2)\ cm$<br>$\theta = (30 \pm 2)°$ | $\Delta\lambda_1 = -2.48\ nm$<br>$\Delta\lambda_2 = -1.45\ nm$<br>$\Leftrightarrow$<br>$R = 4.0\ cm$<br>$\theta = 30°$ |
| #2 (Figure 7b) | $\Delta\lambda_1 = (-1.72 \pm 0.05)\ nm$<br>$\Delta\lambda_2 = (-1.00 \pm 0.05)\ nm$<br>$\Leftrightarrow$<br>$R = (6.2 \pm 0.3)\ cm$<br>$\theta = (30 \pm 3)°$ | $\Delta\lambda_1 = -1.72\ nm$<br>$\Delta\lambda_2 = -1.00\ nm$<br>$\Leftrightarrow$<br>$R = 6.0\ cm$<br>$\theta = 30°$ |
| #3 (Figure 7c) | $\Delta\lambda_1 = (-1.77 \pm 0.05)\ nm$<br>$\Delta\lambda_2 = (0.70 \pm 0.05)\ nm$<br>$\Leftrightarrow$<br>$R = (5.8 \pm 0.2)\ cm$<br>$\theta = (-28 \pm 3)°$ | $\Delta\lambda_1 = -1.77\ nm$<br>$\Delta\lambda_2 = 0.70\ nm$<br>$\Leftrightarrow$<br>$R = 6.0\ cm$<br>$\theta = -30°$ |
| #4 (Figure 7d) | $\Delta\lambda_1 = (-0.84 \pm 0.05)\ nm$<br>$\Delta\lambda_2 = (-0.71 \pm 0.05)\ nm$<br>$\Leftrightarrow$<br>$R = (12.7 \pm 0.9)\ cm$<br>$\theta = (45 \pm 5)°$ | $\Delta\lambda_1 = -0.84\ nm$<br>$\Delta\lambda_2 = -0.71\ nm$<br>$\Leftrightarrow$<br>$R = 12.0\ cm$<br>$\theta = 45°$ |

## 5. Conclusions

In this manuscript, we proposed and demonstrated a new HCPCF-based curvature sensor able to resolve the bending radius and direction. The sensor employs a SR-TL HCPCF displaying tubes with three different thicknesses which are judiciously arranged on the fiber cross-section to provide a bend-dependent optical response. Indeed, by positioning the tubes with different thicknesses in orthogonal directions within the fiber microstructure, one can attain curvature-dependent shifts on the tubes resonance's spectral positions. Thus, via a convenient calibration procedure, which encompasses suitable fitting of empirical functions to the sensor optical response, the curvature radii and angles can be adequately retrieved. We understand that the sensor proposed herein provides new insight on the utilization of hollow-core fibers to attain angle-resolved curvature sensors.

**Author Contributions:** Conceptualization, C.M.B.C., and J.H.O.; methodology, C.M.B.C., M.A.R.F. and J.H.O.; numerical simulations, W.M.G. and M.A.R.F.; validation, C.M.B.C., M.A.R.F., and J.H.O.; data curation, W.M.G., and J.H.O..; writing—original draft preparation, J.H.O.; writing—review and editing, W.M.G., C.M.B.C., M.A.R.F. and J.H.O.; supervision, C.M.B.C., M.A.R.F., and J.H.O.; All authors have read and agreed to the published version of the manuscript.

**Data Availability Statement:** Data underlying the results reported herein can be requested from the authors upon reasonable request.

**Acknowledgments:** C. M. B. Cordeiro and M. A. R. Franco thank the National Council for Scientific and Technological Development (CNPq).

**Conflicts of Interest:** The authors declare no conflict of interest.

## References

1. Debord, B., Amrani, F., Vincetti, L., Gérôme, F., Benabid, F. Hollow-core fiber technology: the rising of 'gas photonics'. *Fibers*. **2019**, 7, 16.
2. Gao, S-F., Wang, Y-Y., Ding, W., Jiang, D-L., Gu, S., Zhang, X., Wang, P. Hollow-core conjoined-tube negative-curvature fibre with ultralow loss. *Nat. Commun*. **2018**, 2828.
3. Amrani, F., Osório, J. H., Delahaye, F., Giovanardi, F., Vincetti, L., Debord, B., Gérôme, F., Benabid, F. Low-loss single-mode hybrid-lattice hollow-core photonic-crystal fibre. *Light Sci. Appl*. **2021**, 10, 7.
4. Sakr, H., Chen, Y., Jasion, G. T., Bradley, T. D., Hayes, J. R., Mulvad, H. C. H., Davidson, I. A., Fokoua, E. N., Poletti, F. Hollow core optical fibres with comparable attenuation to silica fibres between 600 and 1100 nm. *Nat. Commun*. **2020**, 11, 6030.
5. Chafer, M., Osório, J. H., Amrani, F., Delahaye, F., Maurel, M., Debord, B., Gérôme, F. benabid, F. 1-km hollow-core fiber with loss at the silica Rayleigh limit in the green spectral range. *IEEE Photon. Technol. Lett*. **2019**, 31, 9, 685-688.
6. Gao, S-F., Wang, Y-Y., Ding, W., Hong, Y-F., Wang, P. Conquering the Rayleigh scattering limit of silica glass fiber at visible wavelengths with a hollow-core fiber approach. *Laser Photonics Rev.* **2019**, 14, 1, 1900241.
7. Osório, J. H., Amrani, F., Delahaye, F., Dhaybi, A., Vasko, K., Tessier, G., Giovanardi, F., Vincetti, L., Debord, B., Gérôme, F., Benabid, F. Hollow-core fiber with ultralow loss in the ultraviolet range and sub-thermodynamic equilibrium surface-roughness. arXiv:2105.11900, **2021**.
8. Cordier, M., Delaye, P., Gérôme, F., Benabid, F., Zaquine, I. Raman-free fibered photon-pair source. *Sci. Rep.* **2020**. 10, 1650.
9. Okaba, S., Yu, D., Vincetti, L., Benabid, F., Katori, H. Superradiance from lattice-confined atoms inside hollow core fibre. Commun. Phys. **2019**, 2, 136.
10. Chafer, M., Osório, J. H., Dhaybi, A., Ravetta, F., Amrani, F., Delahaye, F., Debord, B., Caiteau-Fischbach, C., Ancellet, G., Gérôme, F., Benabid, F. Near- and middle-ultraviolet reconfigurable Raman source using a record-low UV/visible transmission loss inhibited-coupling hollow-core fiber. arXiv:2108.11327, **2021**.
11. Yu, R., Chen, Y., Shui, L., Xiao, L. Hollow-core photonic crystal fiber gas sensing. *Sensors*. **2020**, 20, 2996.

12. Cubillas, A. M., Jiang, X., Euser, T. G., Taccardi, N., Etzold, B. J. M., Wasserscheid, P., Russell, P. St. J. Photochemistry in a soft-glass single-ring hollow-core photonic crystal fiber. *Analyst*. **2017**, 142, 925-929.
13. Nissen, M., Doherty, B., Hamperl, J., Kobelke, J., Weber, K., Henkel, T., Schmidt, M. A. UV absorption spectroscopy in water-filled antiresonant hollow core fibers for pharmaceutical detection. *Sensors*. **2018**, 18, 478.
14. Sardar, M. R., Faisal, M., Ahmed, K. Simple hollow core photonic crystal fiber for monitoring carbon dioxide gas with very high accuracy. *Sens. Bio-Sens. Res.* **2021**, 31, 100401.
15. Arman, H., Olyaee, S. Realization of low confinement loss acetylene gas sensor by using hollow-core photonic bandgap fiber. *Opt. Quantum Electron*. **2021**, 53, 328.
16. Sardar, M. R., Faisal, M., Ahmed, K. Design and characterization of rectangular slotted porous core photonic crystal fiber for sensing $CO_2$ gas. *Sens. Bio-Sens. Res.* **2021**, 30, 100379.
17. Ravaille, A., Feugnet, G., Debord, B., Gérôme, F., Benabid, F., Bretenaker, F. Rotation measurements using a resonant fiber optic gyroscope based on Kagome fiber. *Appl. Opt.* **2019**, 58, 2198-2204.
18. Sanders, G. A., Taranta, A. A., Narayanan, C., Fokoua, E. N., Mousavi, S. A., Strandjord, L. K., Smiciklas, M., Bradley, T. D., Hayes, J., Jasion, G. T., Qiu, T., Williams, W., Poletti, F., Payne, D. N., "Hollow-core resonator fiber optic gyroscope using nodeless anti-resonant fiber. *Opt. Lett.* **2021**, 46, 46-49.
19. Yang, F., Gyger, F., Thévenaz, L. Intense Brillouing amplification in gas using hollow-core waveguides. *Nat. Photonics*. **2020**, 14, 700-708.
20. Cordeiro, C. M. B., Osório, J. H., Guimarães, W. M., Franco, M. A. R. Azimuthally asymmetric tubular lattice hollow-core optical fiber. *J. Opt. Soc. Am. B*. **2021**, 38, F23-F28.
21. Osório, J. H., Oliveira, R., Aristilde, S., Chesini, G., Franco, M. A. R., Nogueira, R. N., Cordeiro, C. M. B. Bragg gratings in surface-core fibers: refractive index and directional curvature sensing. *Opt. Fiber Technol*. **2017**, 34, 86-90.
22. Newkirk, A. V., Antonio-Lopez, J. E., Valazquez-Benitez, A., Albert, J., Amezcua-Correa, R., Schülzgen, A. Bending sensor combining multicore fiber with a mode-selective photonic lantern. *Opt. Lett*. **2015**, 40, 5188-5191.
23. Mousavi, S. A., Sandoghchi, S. R., Richardson, D. J., Poletti, F., Broadband high birefringence and polarizing hollow core antiresonant fibers. *Opt. Express*. **2016**, 24, 22943-22958.
24. Yerolatsitis, S., Shurvinton, R., Song, P., Zhang, Y., Francis-Jones, R. J. A., Rusimova, K. R. Birefringent anti-resonant hollow-core fiber. *J. Lightwave Technol.* **2020**, 38, 5157-5162.
25. Osório, J. H., Chafer, M., Debord, B., Giovanardi, F., Cordier, M., Maurel, M., Delahaye, F., Amrani, F., Vincetti, L., Gérôme, F., Benabid, F. Tailoring modal properties of inhibited-coupling guiding fibers by cladding modification. *Sci. Rep.* **2019**, 9, 1376.
26. Wang, C., Yu, R., Debord, B., Gérôme, F., Benabid, F., Chiang, K. S., Xiao, L. Ultralow-loss fusion splicing between negative curvature hollow-core fibers and conventional SMFs with a reverse-tapering method. *Opt. Express*. **2021**, 22470-22478.
27. Suslov, D., Komanec, M., Fokoua, E. R. N., Dousek, D., Zhong, A., Zvánovec, S., Bradley, T. D., Poletti, F., Richardson, D. J., Slavík, R. Low loss and high performance interconnection between standard single-mode fiber and antiresonant hollow-core fiber. *Sci. Rep.* **2021**, 11, 8799.
28. Couny, F., Benabid, F., Roberts, P. J., Light, P. S., Raymer, M. G. Generation and photonic guidance of multi-octave optical frequency combs. *Science*. **2007**, 318, 1118-1121.
29. Debord, B., Amsanpally, A., Chafer, M., Baz, A., Maurel, M., Blondy, J, M., Hugonnot, E., Scol, F., Vincetti, L., Gérôme, F., Benabid, F. Ultralow transmission loss in inhibited-coupling guiding hollow fibers. *Optica.* **2017**, 4, 209-217.
30. Malitson, I. H. Intraspecimen comparison of the refractive index of fused silica. *J. Opt. Soc. Am.* **1965**, 55, 1205-1208.
31. Sakr, H., Bradley, T. D., Jasion, G. T., Fokoua, E. N., Sandoghchi, S. R., Davidson, I. A., Taranta, A., Guerra, G., Shere, W., Chen, Y., Hayes, J. R., Richardson, D. J., Poletti, F. Hollow core NANFs with five nested tubes and record low loss at 850, 1060, 1300 and 1625 nm. *Optical Fiber Communications Conference (OFS)*. **2021**, F3A.4.
32. Wang, K., Dong, X., Köhler, M. H., Kienle, P., Bian, Q., Jakobi, M., Koch, A. W. Advances in optical fiber sensors based on multimode interference (MMI): a review. *IEEE Sens. J.* **2021**, 132-142.
33. Osório, J. H., Guimarães, W. M., Peng, L., Franco, M. A. R., Warren-Smith, S. C., Ebendorff-Heidepriem, H., Cordeiro, C. M. B. Exposed-core fiber multimode interference sensor. *Results Opt.* **2021**, 5, 100125.
34. Setti, V., Vincetti, L., Argyros, A. Flexible tube lattice fibers for terahertz applications. *Opt. Express*. **2013**, 21, 3388-3399.

35. Frosz, M. H., Roth, P., Günendi, M. C., Russell, P. St. J. Analytical formulation for bending loss in single-ring hollow-core photonic crystal fibers. *Photon. Res*. **2017**, 5, 88-91.